\documentclass[aps,preprint,groupeaddress]{revtex4}

\usepackage{amsmath} 
\usepackage{amssymb} 
\usepackage{amsfonts}
\usepackage{biophysj}
\usepackage[dvips]{graphicx} 
\usepackage[]{epsfig} 

\begin{document}
\bibliographystyle{biophysj} 

\title{Sequence-specific size, structure, and stability of tight protein knots} 
\author{Joachim Dzubiella} 
\affiliation{Physics Department T37, Technical University  Munich, 85748 Garching, Germany}
\thanks{To whom correspondence should be addressed. E-mail: jdzubiel@ph.tum.de}

\singlespacing 

\begin{abstract}
Approximately 1\% of the known protein structures display knotted configurations in their native fold but their function is not understood.  It has been speculated that the entanglement may inhibit mechanical protein unfolding or transport, e.g., as in cellular threading or translocation processes through narrow biological pores.  Protein knot
 manipulation has become accessible in single molecule experiments, e.g.,  leading to
 knot tightening and localization. Here  we investigate {\it tight} peptide knot (TPK) characteristics in detail by pulling selected 3$_{1}$ and 4$_{1}$-knotted peptides using all-atom molecular dynamics computer simulations. We find that the 3$_{1}$ and 4$_{1}$-TPK lengths are typically $\Delta l\simeq 4.7$~nm and $6.9$~nm,  respectively,  for a wide range of tensions ($F\lesssim 1.5$~nN), pointing to a pore diameter of   $\simeq 2$~nm  below which a translocated knotted protein might get stuck.  The 4$_{1}$-knot length is in agreement with recent AFM pulling experiments. Detailed TPK characteristics however, may be {\it sequence-specific}: we find a different size and structural 
behavior in polyglycines, and, strikingly, a strong hydrogen bonding and water molecule trapping 
capability of hydrophobic TPKs due to side chain shielding of the polar TPK core. Water capturing and release is found to be controllable by the tightening force in a few cases. These mechanisms result into a sequence-specific 'locking' and metastability of TPKs  what might lead to a blocking of knotted peptide transport at designated sequence-positions.  Intriguingly, macroscopic tight 4$_{1}$-knot  structures are reproduced microscopically ('figure-of-eight' vs. the  'pretzel') and can be tuned by sequence in contrast to mathematical predictions. Our findings may explain a function of knots in native proteins, challenge previous studies on macromolecular knots, and may find use in bio- and nanotechnology. 

\vspace{0.25cm}

  \end{abstract}

%\keywords{protein knots | molecular dynamics simulation | buried water}

\maketitle

 \section{Introduction}

After the discovery of the first knotted structure in the
native fold of a protein in 1994 \cite{mansfield:nature}, additional
studies \cite{taylor:nature,lua:plos}, and in particular a recent
survey identified almost three hundreds of further knotted proteins,
constituting around 1\% of known structures in the protein data base
\cite{virnau:plos}. Most of them have the simplest 3$_1$ (trefoil)
topology, only a few have been found to posses the more complicated
4$_1$ (figure-eight) and 5$_2$-types of prime knots. While the question of
physiological relevance is still a matter of debate
\cite{taylor,lua:plos, alam} it has been proposed that the entangled
structure might have profound effects on protein folding and (forced or mechanical) unfolding~\cite{alam, virnau:plos, wallin, prakash}, e.g., to inhibit translocation through biological membranes 
or to protect against degradation by proteasome threading. 
In this respect it is tempting  to speculate that the steric blocking of narrow pathways by a 
localized or tightly pulled protein knot may have bio(techno)logical significance. 
Also relevant, cyclotides, a family of proteins that have a cyclic peptide 
backbone, strong biological activity, and high pharmaceutical 
potential, feature a tightly packed cystine knot in their interior~\cite{craik}.
The synthesis and design of artificially interlocked 
molecules has become possible in supramolecular chemistry with applications in bio- or 
nanotechnology, e.g., as molecular receptors, locks, or machines~\cite{interlocked,suitanes}. 
 
The study of tight knot characteristics in (bio)polymers
has a long history of interest as knots easily self-tie and localize in any long chain
\cite{katritch,virnau:jacs,  belmonte,kardar:review}. More than 20 years ago de Gennes argued 
that knots may self-tie in crystallizing or sheared polymer
melts changing their macroscopic relaxation behavior \cite{degennes}. 
Possible self-tying mechanisms may be based on electrostatic repulsion~\cite{dommersnes},
entropic tightening in worm-like chains \cite{grosberg}, or
localization of (flat) polymer knots either in confinement \cite{ralf1:prl,ralf2:prl, marcone} 
or in bad solvent conditions \cite{virnau:jacs}.  Externally controlled manipulation and characterization 
of microscopic knots has become accessible experimentally by employing optical tweezer
methods \cite{arai:nature,bao:prl}, or atomic force microscopy (AFM) \cite{alam, hugel:review}. 
From a theoretical perspective, scaling arguments
\cite{grosberg},  (water-free) quantum calculations \cite{saitta:nature}, 
or coarse computer simulations
\cite{virnau:jacs,mansfield, farago:epl, arteca1, alex:biophys,
  huang:jpca} have been used. Almost exclusively, previous studies focused 
on homogeneous systems such as polyethylene, DNA, or actin filaments; only recently 
tight protein knots (TPKs) with specific, inhomogeneous sequences have been investigated by Su{\l}kowska 
{\it et al.} using an (implicit solvent) Go-model \cite{cieplak:prl}.  Due to local geometry 
(e.g., side chain size or kinks in the peptide) and in strong 
contrast to previous findings \cite{dommersnes, bao:prl, alex:biophys,  huang:jpca, metzler} 
knot localization and diffusion was dominated by jumps of 
the knot's ends to specific peptide locations suggesting qualitatively different, 
sequence-dependent fluctuations of TPKs when 
compared to homopolymers. 

Although highly relevant for transport, translocation, and threading processes of knotted proteins 
the size of a TPK has not been determined before. For an educated guess, consider a rope of (contour) length $l_{c}$, 
tie a knot in it and pull it tight. The end-to-end distance $l$ is now 
reduced by $\Delta l= l_{c} - l$ which we refer to in the following 
as the {\it tight knot length}, i.e., the length of the rope involved in the
'open' knot \cite{open}. By dividing $\Delta l$ by the rope thickness $D$, we obtain the
characteristic quantity \cite{pieranski} 
\begin{eqnarray}
\Lambda=\frac{\Delta l}{D} 
\label{lambda}
\end{eqnarray} 
which is minimized by the tight knot conformation, and is 
$\Lambda$=10.1 and 13.7 for a 3$_{1}$ and $4_1$, respectively, 
for idealized hard-core ropes \cite{pieranski}. Assuming now a typical
peptide thickness of the order of an atomic size,
$D\simeq 0.35-0.5$~nm, we expect $\Delta l=\Lambda D\simeq 3.5 - 5.1$~nm~ and $\simeq 4.8 - 6.9$~nm~ for
tight 3$_{1}$ and $4_1$ peptide knots, respectively. If, naively, the knot is further assumed to 
be a circle with diameter $\Delta l=2\pi R$ we infer that a typical TPK radius may be of $R \simeq 0.6-1.1$~nm. 

In this work, we take a more detailed look at TPKs by performing {\it explicit-water} molecular dynamics (MD) computer simulations \cite{andy} of 3$_{1}$ 
and 4$_1$-knots in selected polypeptides involving up to 30 amino acids and systematically study their size and 
structural behavior. We find tight knot lengths $\Delta l \simeq 4.7\pm0.3$~nm (involving $13\pm 1$ amino acids) and
$\simeq 6.9\pm0.3$~nm (19$\pm 1$ amino acids) for the 3$_{1}$ and 4$_{1}$,  
respectively, surprisingly constant for a wide range of stretching forces ($F\lesssim 1.5$~nN), and typical tight 
knot radii of gyration of $R_{g}\simeq 0.7-0.8$ nm, all in the range of the macroscopic estimate~(1). The 4$_{1}$-TPK length is in agreement with recent AFM pulling experiments on the natively knotted bacterial phytochrome~\cite{bornschl}.  Detailed tight knot characteristics however, may be sequence specific,  e.g., we find smaller knots, 
and different structural and stability behavior in the special case of polyglycines. Strikingly, TPKs 
have a strong water capturing and hydrogen bonding capability within their closely packed interior, which 
is sequence-specific and promoted by nonpolar side-chains. Buried water and long-lived intra-knot hydrogen bonds lead to surprisingly rigid and stable tight knots in free simulations on a $\sim 100$ ns time scale.  Intriguingly, macroscopic tight 4$_{1}$-knot  structures are reproduced microscopically ('figure-of-eight' vs. a  'pretzel'-like configuration) but depend on peptide sequence in contrast to mathematical predictions of the tight 4$_{1}$-knot structure which is the 'figure-of-eight' figure~\cite{pieranski}. We predict strongly localized tight knots after peptide stretching and a preferential  affinity towards regions with dominantly nonpolar side chains. We demonstrate that the accurate modeling of specific side chains {\it and} the aqueous environment is  crucial for the full understanding of TPK characteristics. 
  
\section{Methods and systems}

\subsection{MD simulations}

Our all-atom MD simulations are performed using the software package Amber9.0 
with the ff03 force-field  and TIP3P solvent~\cite{amber}. 
Systems are maintained at a fixed pressure of $P=1$ bar and
temperature $T=300$ K by coupling to a Berendsen barostat and
Langevin thermostat, respectively. System sizes vary
between $N\simeq4000$ and $N\simeq8000$ atoms.
Electrostatic neutrality is assured by additional Na$^{+}$-counterions compensating the net peptide
charge given at pH=7.  The rectangular and periodically repeated
simulation box has edge lengths $L_x\simeq L_y\simeq 30$~\AA, while in
peptide stretching direction $L_z\simeq 5.5-7.0$~\AA. Electrostatic
interactions are calculated by particle mesh Ewald summation and
real-space interactions have a cut-off of 9~\AA.
 Polypeptides are generated  using the Amber {\it tleap} tool. Knots are
tied into them utilizing interactive MD (IMD) in VMD~\cite{vmd}: while a Langevin simulation of the peptide is running and 
visualized a force can be applied to selected fragments by using the computer 
mouse so that the peptide can be dragged by hand into a finally knotted 
configuration. Thereafter, the system is equilibrated for $\simeq 5$ ns 
with Langevin dynamics, solvated with TIP3P water, and further 
equilibrated by a $\simeq 5$ ns MD simulation. For peptide stretching and loosening
we utilize  the Amber steered MD (SMD) tool: a constant pulling 
velocity of 0.1~\AA/ns (0.01 m/s) drives the first and last atom (in a distance $l$) 
of the peptide backbone in opposite directions and force-extension curves $F(l)$ 
are calculated. Pulling is terminated after the mean force reaches $\sim $1.5~nN, a value 
at which covalent bond breaking can occur experimentally \cite{hugel:review}. 
Simulation 
snapshots are generated using VMD~\cite{vmd}. Hydrogen bonds, 
radii of gyration, and rms deviations are  analyzed using the Amber {\it ptraj} tool.
 
\subsection{Systems}

$3_{1}$ and $4_{1}$-types of knots are investigated. To study the influence of amino 
acid type on the tight knot structure we opt for three different homopeptides: the hydrophobic polyleucine 
(sequence L$_{N_{\rm aa}}$), the 
partly hydrophilic and charged polyglutamic acid (E$_{N_{\rm aa}}$), and the 
slim, amphiphilic polyglycine (G$_{N_{\rm aa}}$).  The peptides have a total number 
of $N_{\rm aa}$=21 and 30 amino acids for the $3_{1}$ and $4_1$-knots, respectively. 
Furthermore, two randomly
picked pieces from the knotted cores of the natively $3_1$-knotted
YibK methyltransferase \cite{lim}
and the $4_{1}$-knotted Class II ketol-acid reductoisomerase \cite{biou} 
are considered to directly connect to naturally occurring protein knots. 
In the following we name the knotted peptides by knot type and sequence, 
e.g., '$3_{1}$L' for a polyleucine trefoil and '$4_{1}$mix' for the 
$4_{1}$-knot in a mixed sequence. The different knotted peptide systems, their 
amino acid (aa)  sequences and numbers $N_{\rm aa}$ are summarized in Tab.~I.

\section{Results and discussion}

\subsection{Tight knot size and structure}

A typical initial configuration of a 3$_{1}$-knotted peptide in our
simulation is shown in Fig.~1~a) where a snapshot of 3$_1$G is 
sketched before pulling it tight. The end-to-end extension here is $l\simeq 25$\AA. 
A tight knot situation for the same peptide is shown in Fig.~1~b) for a large stretching 
force of $\sim$1.5~nN ($l\simeq 45$\AA).  For an elastic peptide as considered in this study the final
'tightness' of the knot will naturally depend on the external
stretching force $F$. The calculated force-extension curve, $F(l)$, for
 3$_1$G is shown in Fig.~2 a) together with the data for 3$_{1}$E and 4$_{1}$L.  
We observe an overall monotonic nonlinear increase of the force.  Fluctuations are 
moderate on that scale and have local standard deviations ranging 
from $\sim 20$ pN up to $\sim 50$ pN. Note that we also plot $F(l)$ of knot 
loosening, i.e., 'reverse pulling', showing no obvious hysteresis. 
This indicates that our systems are close to equilibrium at 
the chosen pulling rate of 0.1~\AA/ns.

In order to determine the tight knot length $\Delta l(F)=l_c(F)-l(F)$
an accurate estimate for the force-dependent contour length
$l_{c}(F)$ of the unknotted peptide is needed. For this, we calculate
the average amino acid length $\Delta
l_{\rm aa}(F)$ by measuring the mean distance  between 
neighboring backbone nitrogen atoms in short unknotted peptides, presented 
in the inset to Fig.~2 a): below a stretching force of around $\sim$10 pN
the length thermally fluctuates around $\Delta l_{\rm
  aa}(F)\simeq$ 3.5~\AA, then rises quickly with
force in a nonlinear fashion in the low-stretching, thermal regime
($F\sim$ 10 - 150 pN) to eventually increase linearly in the high
stretching regime $F\gtrsim$ 150~pN.  At $F = $ 1.5~nN a value 
of $\Delta l_{\rm aa}(F)\simeq 3.8$~\AA~ is
reached. From the slope $b$ of the linear part we estimate the linear
elastic modulus $\Gamma =\Delta l_{\rm aa}(F=0)/b \simeq 42$~nN, 
which is in agreement with AFM pulling experiments, 
where $\Gamma\simeq 50\pm15$~nN \cite{ptak}. This agreement is 
remarkable since MD force-fields are typically not benchmarked to 
be accurate at considered large tensions.

In pulling experiments, rupture of some terminal bonds at the AFM tip
can occur at forces of around $F_{1}=$200~pN \cite{hugel:review},
where we find $\Delta l_{\rm aa}(F_1)\simeq 3.68$~\AA, leading to contour length
estimates $l_c(F_{1})=N_{\rm aa}\Delta l_{\rm aa}\simeq77.3$~\AA~ and $l_c(F_{1})\simeq 110.4$~\AA~ for the trefoil and the $4_1$
peptides, respectively. Consequently, it follows that the tight knot lengths for
the trefoil peptides  are between $\Delta l(F_1)\simeq 44.3$~\AA~ (3$_1$G)
 and $\simeq $ 49.8~\AA~ (3$_{1}$L). The number of amino 
acids involved in the knot are thus $n_{\rm aa}=\Delta l/\Delta l_{\rm aa}\simeq 13$.
For the $4_1$-knots  three of the four values lie between $\Delta l\simeq $ 68.4~\AA~ and 
71.9~\AA~ ($n_{\rm aa}\simeq 19$) while for the polyglycine knot (4$_1$G) we find 
$\Delta l(F_{1})\simeq 60.4$~\AA~ ($n_{\rm aa}\simeq 13$), around 14\% smaller. The
lengths are summarized in Tab.~I.  A typical error of these values is
given by the fluctuations of the $F(l)$ curve and is roughly of amino
acid size ($\pm 3$~\AA).

Let us know consider more intense stretching and study the knot
lengths at a larger force $F_{2}=1$~nN. $\Delta l_{\rm aa}(F)$
increases to $\simeq$ 3.76~\AA~ giving rise to a slightly larger contour
length for the unknotted peptides.  Evaluating the
particular knot lengths we observe that the knots shrink in
size (while the whole peptide is more stretched) as could have been
anticipated. Although the pulling force is substantially increased 
typically only one amino acid less is involved in a single knot, 
so that surprisingly the knot sizes vary only a few percent for a wide range 
of tensions. An exception however, are both of the polyglycine peptides: here the
tightening effect is considerable and the final knot lengths are 20-25\%
smaller than those of the other studied peptides. All lengths are
summarized in Tab.~I.

The knot lengths of $\simeq 4.7$~nm and $\simeq 6.9$~nm for the 3$_{1}$ and 4$_{1}$-knots, respectively, fall inside the range of prediction (\ref{lambda}) indicating 
on a first glance that they are primarily  determined by {\it generic} packing effects with an effective excluded volume thickness $D$ similar for most of the peptides. Furthermore,  
macroscopic arguments roughly hold on the molecular scale. 
In contrast, hydrophobicity and hydrophilicity seem to have no direct influence 
on tight knot size in the considered force regime.  Examining a
bit closer the nature of the amino acid side chains  supports this statement:
while the glycine side chain is identical to just a single hydrogen atom, 
a typical residue with a few carbon atoms gives rise to a more difficult molecular 
arrangement close to or inside the tight knot. This  presumably leads to the
20-25\% smaller knots in the special case of polyglycine.  For the latter we thus
find a smaller effective thickness $D\simeq 3.7$~\AA, while for the
other peptides $D\simeq 4.6-5.0$~\AA. Importantly, apart from the polyglycine, 
all 4$_{1}$-TPK lengths are in agreement with recent AFM pulling experiments on the natively knotted bacterial phytochrome~\cite{bornschl}.

Illustrating simulation snapshots are shown in Fig.~3 where we plot
tight knot situations for the peptides 4$_1$G and 4$_1$mix including
their side chains. Large side chains obviously impede tight peptide
packing.  We also calculate the radius of gyration $R_{g}$ 
of the knots, i.e., by averaging root mean square (rms) atomic distances 
from the geometric center of the atoms involved in the knot, i.e., constituting
the length $\Delta l$.  We measure  
$R_{g}\simeq 7.2\pm0.2$~\AA~  and $R_{g}\simeq 7.8\pm0.2$~\AA~
for the 3$_1$ and 4$_{1}$ knots, respectively,  with only weak dependence 
on the stretching force  for all considered peptides apart from the polyglycine. 
For the latter radii of gyration are found to be close to the values above for weak stretching ($F\sim$ 200 pN)  but 20\% smaller for strong stretching ($F\gtrsim$ 1 nN).
These TPK sizes are larger or comparable as the size of biological 
channels such as in the protease~\cite{prakash,pickart} so that a translocation of a
 knotted protein would indeed be blocked by the presence of a knot.

An intriguing topological feature appears when inspecting the overall 4$_{1}$-knot 
structure  without the obscuring side chains, as in Fig. 1 c) and d). While the 4$_1$G knot
is figuratively indeed in a figure-eight configuration 4$_1$L displays a 
'pretzel'-like configuration.  Actually, we find that {\it all} the
considered $4_1$-knots expect from 4$_1$G prefer the pretzel
when inspected by eye. This comes as a surprise as the tight 4$_{1}$-knot 
configuration which minimizes (1) has been shown to be figuratively 
the  'figure-of-eight',  at least using simplifying mathematical 
assumptions \cite{pieranski}. Presumably the reasons are rather 
physics-based, i.e., high friction or lowering of the system free energy 
due  more favorable amino acid interactions and packing around the 
knot region may result into the pretzel. Interestingly, the pretzel-like 
configuration can be a  stable 4$_{1}$-configuration in macroscopic knots under 
tension, e.g., as can be easily self-demonstrated using a simple computer cable, or 
as taught in books on cowboy rope tricks, see Fig. 4.

\subsection{Water trapping, hysteresis, and hydrogen bonding}

A striking structural feature we observe is the capability of some peptide 
knots to capture and strongly bind water molecules in their
interior.  The simulated peptides show this effect with varying
magnitude, i.e., we find no bound water in polyglycine (3$_1$G
and 4$_1$G) and the mixed peptide 3$_1$mix for any simulated peptide
extension, while in 3$_1$E a single trapped water molecule
is reproducibly found only in the case of very close peptide packing at high
forces $F\gtrsim $1 nN.  We find stronger water binding qualities for
the other four peptides 3$_1$L, 4$_1$E,  4$_1$L, 4$_{1}$mix for a 
wider range of simulated peptide extensions, i.e., water was bound for simulation
times of the order of $\sim$10-100~ns per peptide pointing to a quite
stable mechanism. On a first glance surprisingly, both homopeptides
with the purely hydrophobic leucine side chains show the strongest
water trapping capability.

Simulation snapshots are shown in Fig.~5 for the peptides 4$_1$E and
4$_1$mix: the water bonds to the backbone amides in the 
knot interior, involving at least three hydrogen bonds per molecule, and is 
rotationally immobilized. Apparently the water binding is made 
possible by the tight peptide packing in the highly bent knot allowing for multiple
bonds of a water molecule to the polar backbone. A particularly
interesting case is the water binding in 3$_1$L. Here the bound 
water molecule is {\it  squeezed out} of the knot interior for large stretching forces
$F\simeq $1 nN.  This behavior leads to a strong force peak in the
force extension curve as shown in Fig.~2 b): for extensions $l\lesssim
30$~\AA~ the water molecule is bound as
shown in the left snapshot.  At $l\simeq 30$~\AA~ and $F\simeq 1$~nN 
the bound water is 'wrung' out and the force drops significantly before further
increasing.  When the knot is loosened $F(l)$ shows a considerable
hysteresis. However, a water molecule is captured by the knot again during 
loosening at extensions $l \lesssim 27$~\AA~ and $F\simeq 200$~pN. 
Repeating the stretching-loosening 
loop twice shows {\it quantitative reproducibility} of this effect [cf. Fig.~2 b)]. 
The occurrence of the hysteresis points to the fact that the water binding-unbinding
events fluctuate on large time-scales and this simulation
deviates therefore from equilibrium. The magnitude of the hysteresis
can be estimated by integrating over the $F(l)$ stretching-loosening
cycle which gives rise to a large dissipation energy of about $\Delta
G\simeq 30-35~k_{\rm B}T$, indeed comparable to the energy of 3-4 hydrogen
bonds between a water molecule and a peptide environment (8-10 $k_{\rm B}T$
per hydrogen bond) \cite{jackson:bio}.  

It is a well-known fact that buried water molecules constitute an integral part 
of many native protein structures contributing to stability, flexibility, folding, and
mechanical and enzymatic function \cite{baker, denisov:nature, dougan, ball:review}. 
Noteworthy, our measured $\Delta G$ is very close the binding enthalpy of a buried water
molecule in the polar pocket of bovine pancreatic trypsin inhibitor
(BPTI) \cite{fischer:pnas}, where four hydrogen-bonds constitute $
\Delta H\simeq 36~k_{\rm B}T$. We find a similar large dissipation energy in
peptide 4$_{1}$mix ($\Delta G \simeq 20 k_{B}T$) and less pronounced in 
4$_{1}$E ($\Delta G \simeq 12 k_{B}T$) and 3$_{1}$E ($\Delta G \simeq 5 k_{B}T$) 
due to partial water hydrogen binding events during knot tightening. No hysteresis 
is found in 4$_{1}$L as water is bound here during the full stretching-loosening 
loop without any binding/unbinding transition. 

Interestingly, in the sequence 3$_{1}$mix we find no water
trapped in the knot interior for all peptide extension in contrast to
4$_{1}$mix, where we observe water bound on a $\sim 10$ ns time scale
with three binding/unbinding events for tension $F\lesssim 500$~pN. 
A closer inspection  of the MD trajectory
reveals that the immediate surrounding of the buried water molecule
consists of 6 amino acids, ALD FQS, which create a mostly hydrophobic
environment, see Fig. 5 b).  This observation and the strong water binding capabilities of
the polyleucines indicate that nonpolar side chain 
environments promote water hydrogen bonding to the tightly  
packed polar backbone. We explain this by the textbook-fact that hydrogen 
bonds are generally stronger in a nonpolar and/or {\it desolvated} protein 
environment \cite{jackson:bio, roseman}, where electrostatic interactions are 
only weakly screened. We suspect that additionally the hydrophobic side chains impose a 
large energy barrier for a possible escape of a water molecule. 
In the polyglutamic acids water screening and the (probably lower) barrier 
is likely to be  provided by the methylene groups of the side chains immediately surrounding the
knotted peptide region. A nonpolar environment is clearly absent in the 
polyglycines. However, strong water trapping capability seems to results 
from a unique and delicate combination of local backbone structure and a 
specific, but rather nonpolar amino acid side chain environment.  

Related to this, another consequence of the tight peptide packing as further 
revealed by our simulations is the existence of long-lived hydrogen bonds between 
particular backbone amide groups.  For instance during the $\simeq 200$ ns 
stretching and loosening loop of polyleucine 4$_{1}$L  we find that amino acids 
10 and 24 hydrogen bond for $\simeq 80\%$ of the simulation time. 
Detailed analysis yields similar behavior for the other peptides, i.e., 
intrapeptide hydrogen bonds are stable on a long $\sim$10-100 ns time scale. 
An exception is polyglycine, where the longest hydrogen bond 
life expectancy is found to be one or two orders of magnitude shorter.

\subsection{Free simulations and tight knot stability}

We also conduct free simulations of the knotted peptides without any
constraints in order to check whether the knots dissolve on a typical
simulation time scale. Initial configurations are taken from 
a stretched situation with $F \simeq 200$ pN. Dissolution of a knot is 
loosely defined here by connecting the peptide ends with an imaginary line and 
looking whether we find a knot in the closed loop or not. Only
both of the polyglycine knots, 3$_{1}$G and 4$_{1}$G, show strong
fluctuations and unknot quickly on a time scale of $\sim$10 ns.  
All other investigated knots do {\it not} dissolve in a $\simeq$ 120 ns 
simulation pointing to a (meta)stable tight knot situation. 
For quantifying this we measure the rms deviation from the initial structure and find values of 
$\sim 2$~\AA~ increasing quickly within $\sim$~10 ns  to $\simeq 7$~\AA~ 
for the polyglycines.  For the other peptides however, the rms value stays at $\simeq 2$~\AA~ for  the total simulation time, supporting the observation that apart from the 
polyglycines tight knots stay stable and quite rigid after peptide stretching 
on relatively long time scales. We note that dissolution of the polyglycine proceeds rather via a knot 'swelling' than a 'slithering' mechanism (where the knot stays tight and diffuses to the end), possibly to relax the highly bent backbone. This needs not to be in contrast to study
\cite{grosberg} where an entropic tightening and slithering was predicted as this might be the
dominant mechanism for somewhat 'looser' tight knots.

As in the constrained case closer inspection of the knot structure 
reveals a few long-lived hydrogen bonds in all stable knots. 
A representative illustration is shown in Fig.~5~c) where we plot a MD snapshot
of 4$_{1}$mix after a $\simeq $ 120~ns free simulation. Four hydrogen
bonds are found between amide backbone groups
right at the knot's ends clearly inhibiting the opening of the knot. Typically, 
we find that these hydrogen bonds persist on average $\simeq 100$ ns 
(80-90\%) of the total free simulation time, even slightly longer for the 
polyleucines.  In 3$_{1}$mix (where no buried water could
be detected) the longest hydrogen-bond life time is shorter ($\simeq$~50-60~ns). 
Remarkably, in 3$_1$L and 4$_{1}$L  additionally one water molecule was trapped  
during the full, unconstrained simulation, constituting a remarkable total of 7-8 long-lived 
hydrogen bonds within the knots. Again the high quantity and 
persistence strength of hydrogen bonds in 3$_{1}$L and 4$_{1}$L must be 
attributed to the desolvated, strongly nonpolar side chain environment of the 
tight knot.

\section{Concluding remarks}

In summary, our MD study of TPKs have revealed some generic, 
but also unexpectedly specific behavior and provoke some interesting 
speculations and future prospects: 

As previously conjectured, the steric blocking of narrow pathways  
by a localized or tightly pulled  knot might be possible {\it in vivo} and also relevant for biotechnological 
purposes. In this light interesting are our findings that TPKs exhibit an 
unexpectedly strong stability and their radii of gyration are all indeed slightly bigger ($\sim$7-8~\AA) at moderate stretching ($F \lesssim 200$ pN) than the radius of the protease pore ($\sim$6.5~\AA)~\cite{prakash,pickart}. We predict that a translocated knotted protein should get stuck in pores with  a diameter below $\simeq 2$~nm.
 
Interesting for further investigation not only from a topological 
point of view \cite{stasiak:nature} is the observation that most 4$_{1}$-knots are figuratively not in a figure-eight but rather in a 'pretzel'-like configuration, which might be a (meta)stable 
configuration in 'physical' open tight knots in contrast to those underlying simplifying
mathematical assumptions \cite{pieranski}.  The pretzel may be stuck by high 
molecular friction, e.g., caused by hydrogen bonds,  and/or preferred by the lowering of the  system free energy due to favorable amino acid  arrangements.   

As a striking result we find that the TPK interior has a strong water binding  
and hydrogen bonding capability which is promoted in rather nonpolar side-chain environments.   These mechanisms result in 'locking' of the  knot structure and surprisingly stable and rigid  tight knots 
after peptide stretching in unconstrained MD simulations on a $\sim 100$~ns time scale. 
The observed {\it quantitative} reproducibility of squeezing-out and capturing a water molecule at  well-defined tensions  may allow for an external {\it mechanical control} of the capturing and releasing  of single water molecules by designed peptide knots. Important in this respect is that buried water is known to be an 
integral part of native protein structures and not only affects protein flexibility and folding~\cite{fischer:pnas, denisov:nature,dougan} but can be essential for catalytic action~\cite{ball:review}. 

Furthermore,  TPKs might resemble structural elements of the cyclotide protein family -- constituted by short ($\sim 30$ aa) cyclic peptides with a tight cystine knot -- which has strong potential in drug design \cite{craik}.  In view of their structural complexity and close relation to native phenomena (engineered) TPKs  thus might serve as an important model system for a  deeper understanding of protein folding and stability \cite{yeates}, and  enzymatic activity, and may be useful for pharmaceutical 
purposes due to a possible catalytic function.

Finally, we would like to encourage further experiments in this stimulating field which are readily available using AFM~\cite{bornschl} or optical tweezer methods~\cite{arai:nature, bao:prl}. Desirable are studies on protein knot length and 
size, stability, and diffusion behavior along stretched peptides and the refolding of knotted proteins
after stretching.  Particularly interesting is knotted peptide translocation for probing the mechanical 
forces involved in protein unfolding, knot tightening and blocking, e.g., by threading them through narrow 
biological or solid-state nanopores \cite{dekker}.  Buried water molecules in peptide knots might be 
detectable by nuclear magnetic relaxation dispersion methods (NMRD) \cite{denisov:nature} allowing for 
another experimental method to explore TPK fluctuations and energy landscapes. 

\begin{acknowledgments}
  J. D. is grateful to Thomas Bornschl\"ogl, Katrina Forest, and Matthias Rief for 
  pointing to this interesting project,  Lyderic Bocquet, Ralf Metzler, Roland Netz, 
  and Joachim Seel for useful comments, and
   the Deutsche Forschungsgemeinschaft (DFG) for support
  within the Emmy-Noether-Program.  Computing time on the HLRBII computer
  cluster of the Leibniz-Rechenzentrum M\"unchen is acknowledged. 
\end{acknowledgments}

%\end{article}

\newpage 

\begin{table}[h]
\begin{center}
\begin{tabular}{|l || c | c || c | c | c | c || c | c | c | c |}
\hline
  Knot  & aa-sequence &$N_{\rm aa}$&$l_{\rm c}(F_{1})$/\AA & $l(F_{1})$/\AA & $\Delta l(F_{1})$/\AA & $n_{\rm aa}(F_{1})$  &$l_{\rm c}(F_{2})$/\AA & $l(F_{2})$/\AA & $\Delta l(F_{2})$/\AA & $n_{\rm aa}(F_{2})$\\  
\hline 
\hline
3$_1$L  & L$_{21}$ &21         & 77.3 & 27.5  & 49.8 &  14  & 79.0 & 32.5  & 46.5 &  12 \\
3$_1$E  & E$_{21}$        &21 & 77.3 & 31.0  & 46.3 &  12 & 79.0 & 32.5  & 46.5 &  12 \\
3$_1$G  & G$_{21}$        &21& 77.3 & 33.0  & 44.3 &  12  &  79.0 & 42.0  & 37.0 &  10  \\
\hline
3$_1$mix &AHSQVKFKLG  &21& 77.3 & 27.8  & 49.5 &  13  & 79.0 & 32.9  & 46.1 &  12\\ 
         &DYLMFGPETRG &  &      &       &      &      &&&&\\
\hline
\hline
4$_1$L  &L$_{30}$ &30& 110.4 & 40.0  & 70.4 &  19  & 112.8 & 44.0  & 68.8 &  18 \\
4$_1$E  &E$_{30}$ &30& 110.4 & 42.0  & 68.4 &  19  & 112.8 & 45.1  & 67.3 &  18 \\
4$_1$G  &G$_{30}$ &30& 110.4 & 50.0  & 60.4 &  16  & 112.8 & 63.0  & 49.8 &  13 \\
\hline
4$_1$mix &TKGMLALYNS &30& 110.4 & 38.7  & 71.7 & 19  & 112.8 & 44.3 & 68.5 & 18 \\ 
    &LSEEGKKDFQ & &  &   &  &    &&&&\\
    &AAYSASYYPS & &  &   &  &    &&&&\\
\hline
\hline
\end{tabular}
  \caption{Simulated knotted peptide systems. The peptides have $N_{\rm aa}$ amino acids (aa) with shown sequence. $l_{\rm c}$ is the estimated contour length of the unknotted peptide, $l$ the measured end-to-end distance of the knotted peptide, and $\Delta l$ the tight knot length involving $n_{\rm aa}$ amino acids. The lengths are evaluated at a stretching force of $F_{1}$=200 pN and $F_{2}$=1 nN.}

\label{tab1}
\end{center}
\end{table}

\newpage 

\begin{figure}[htp]
\caption{MD simulation snapshots of different protein knots in a
  'cartoon' representation. a) Initial configuration of peptide 3$_1$G, 
  where $l\simeq 25$~\AA~ is the end-to-end distance. b) 
  Tight knot configuration of peptide 3$_1$G. The end-to-end distance 
  is $l\simeq 45.0$~\AA~ at a stretching force  $F\simeq$ 1.5~nN. c) 
  A tight 'figure-eight' knot configuration of peptide 4$_1$G at 
   $F\sim$1~nN. d) A tight 'pretzel' knot configuration of peptide 4$_1$L 
    at $F\simeq$1~nN.}
\label{fig:1}
\end{figure}

\begin{figure}[htp]
\caption{a) Force ($F$)-extension ($l$) curves for the peptides 3$_{1}$E,
  3$_{1}$G, and 4$_{1}$L. Stretching curves (solid black lines) and
  loosening curves (dashed red lines) lie on top of each other
  indicating a small hysteresis. Pulling rate is 0.1~\AA/ns.  The
  inset shows the mean distance $\Delta l_{\rm aa}$  between neighboring backbone nitrogen atoms 
  vs. stretching force $F$ in an unknotted peptide. b) Force-extension curves for the polyleucine 3$_1$L.    Black lines correspond to stretching while red lines
  correspond to loosening of the knot. While stretching, for
  extensions $l\lesssim 30$~\AA~ a single water molecule is
  permanently trapped by the polar backbone of the peptide knot, 
  see left snapshot. When exceeding $\simeq 30$~\AA~  the water
  molecule is squeezed out (right snapshot) giving rise to a
  significant peak in the force-extension stretching curve. This
  transition leads to a considerable hysteresis when stretching and
  loosening curves are compared. The effect is reproducible when the
  stretching-loosening loop is repeated (dashed lines).}
\label{fig:2}
\end{figure}

\begin{figure}[htp] 
\caption{MD simulation snapshots of the tightly knotted 4$_1$G (a) and
  4$_1$mix (b). The backbone is sketched in a yellow ribbon for better identification and all amino acids are resolved in a 'licorice' representation ($F\simeq 1$~nN).}
\label{fig:3}
\end{figure}
 
\begin{figure}[htp]
\caption{Cowboy rope trick from B. S. Mason's book \cite{cowboy} showing a 4$_{1}$-knot with a 'pretzel'-like structure
 (left) and a 'figure-of-eight' structure (right). Fascinatingly, we find that both configurations are stable in proteins microscopica
lly and depending on protein sequence. The book says instructively ``The figure-of-eight  is tied by precisely the same movements as t
he pretzel. The factor determining which knot will result is the way the knot is dropped after shaking the rope off your arm. Jerk it 
as it falls and you should have the figure-of-eight. Shake it off gently and the pretzel should result.'' (Taken from http://www.inqui
ry.net/outdoor/spin$\_$rope/trick$\_$knots.htm.)}
\end{figure}

\begin{figure}[htp] 
\caption{MD simulation snapshots of water trapped in peptide 4$_1$E (a)
  and 4$_1$mix (b) at a force $F\simeq$ 1~nN. The backbone is shown in ribbon structure and 
  only those residues are sketched ('licorice' representation) which are actively involved in water
  binding. Water (red and white spheres) is hydrogen bonded to the backbone amides. 
  c) MD snapshot of 4$_{1}$mix in an
  unconstrained MD simulation. Four long-lived hydrogen bonds between backbone 
  amides  at the knot's ends are explicitly drawn (dotted blue lines). }
\label{fig:4}
\end{figure}

\clearpage
\newpage

\begin{figure}[htp]
Fig. 1
\includegraphics[width=9cm,angle=0]{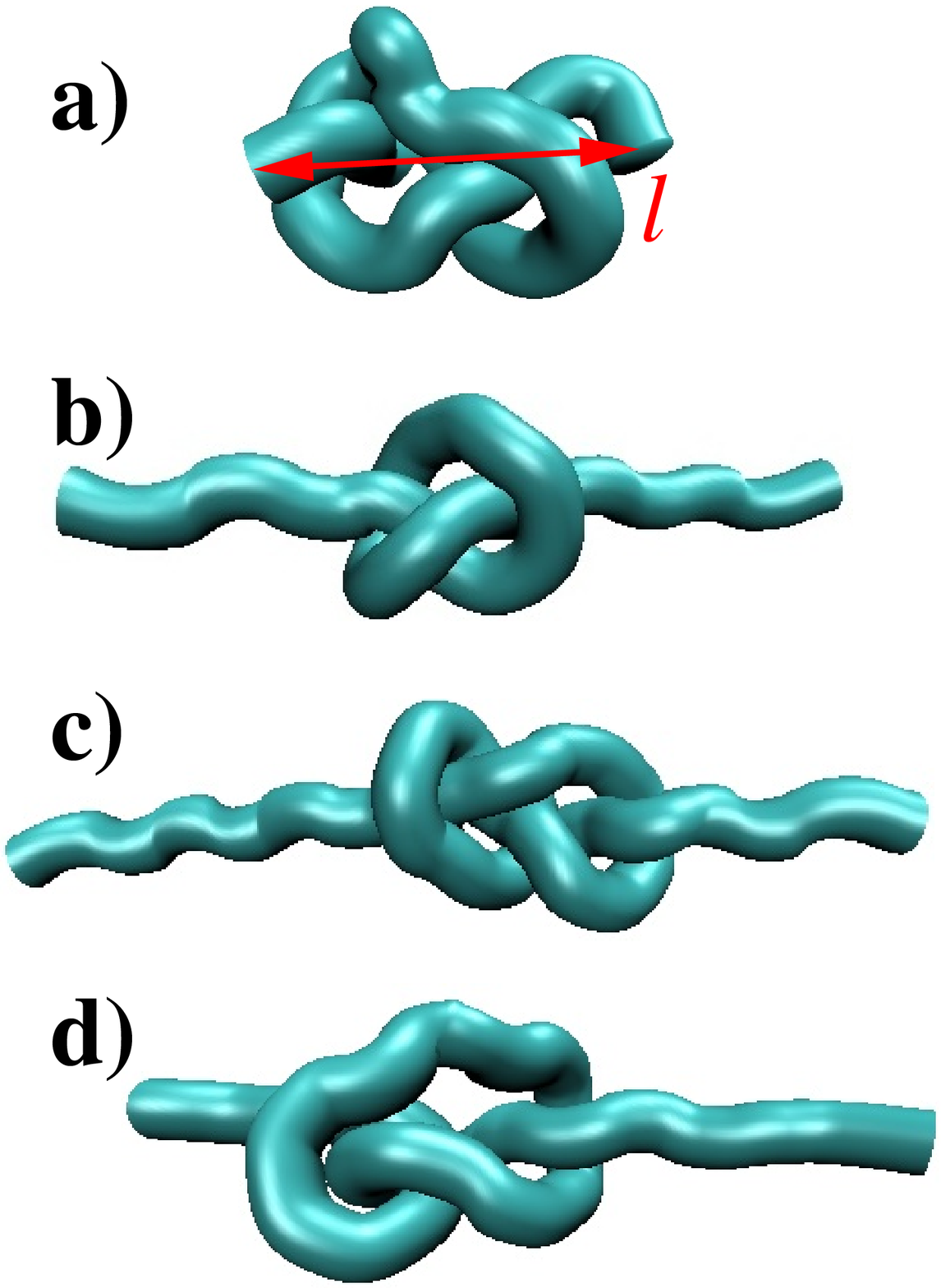}
\label{fig:1}
\end{figure}

\newpage

\begin{figure}[htp]
Fig. 2
\includegraphics[width=12cm,angle=0]{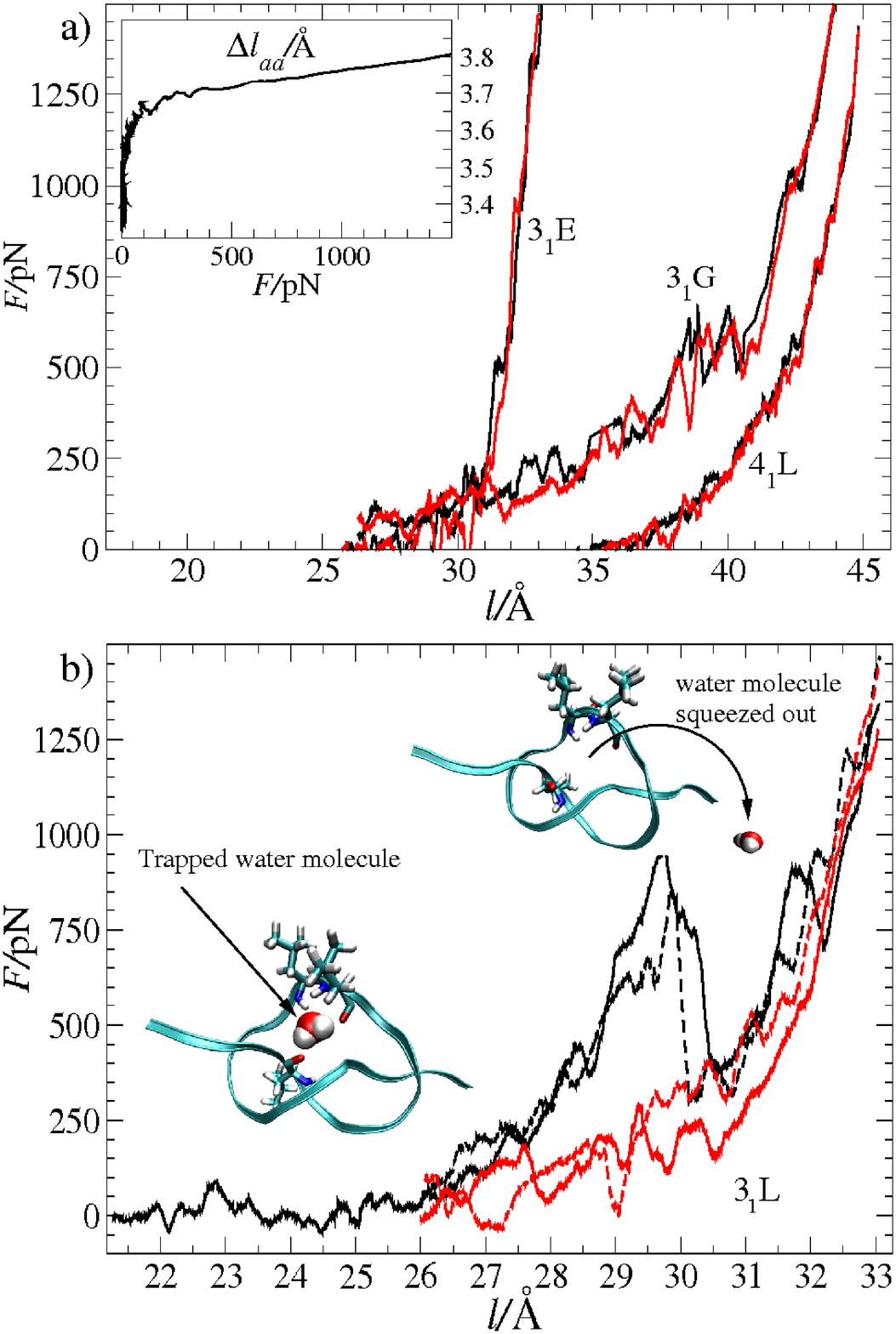}
\label{fig:2}
\end{figure}

\newpage 
\begin{figure}[htp] 
Fig. 3
\includegraphics[width=14cm,angle=0]{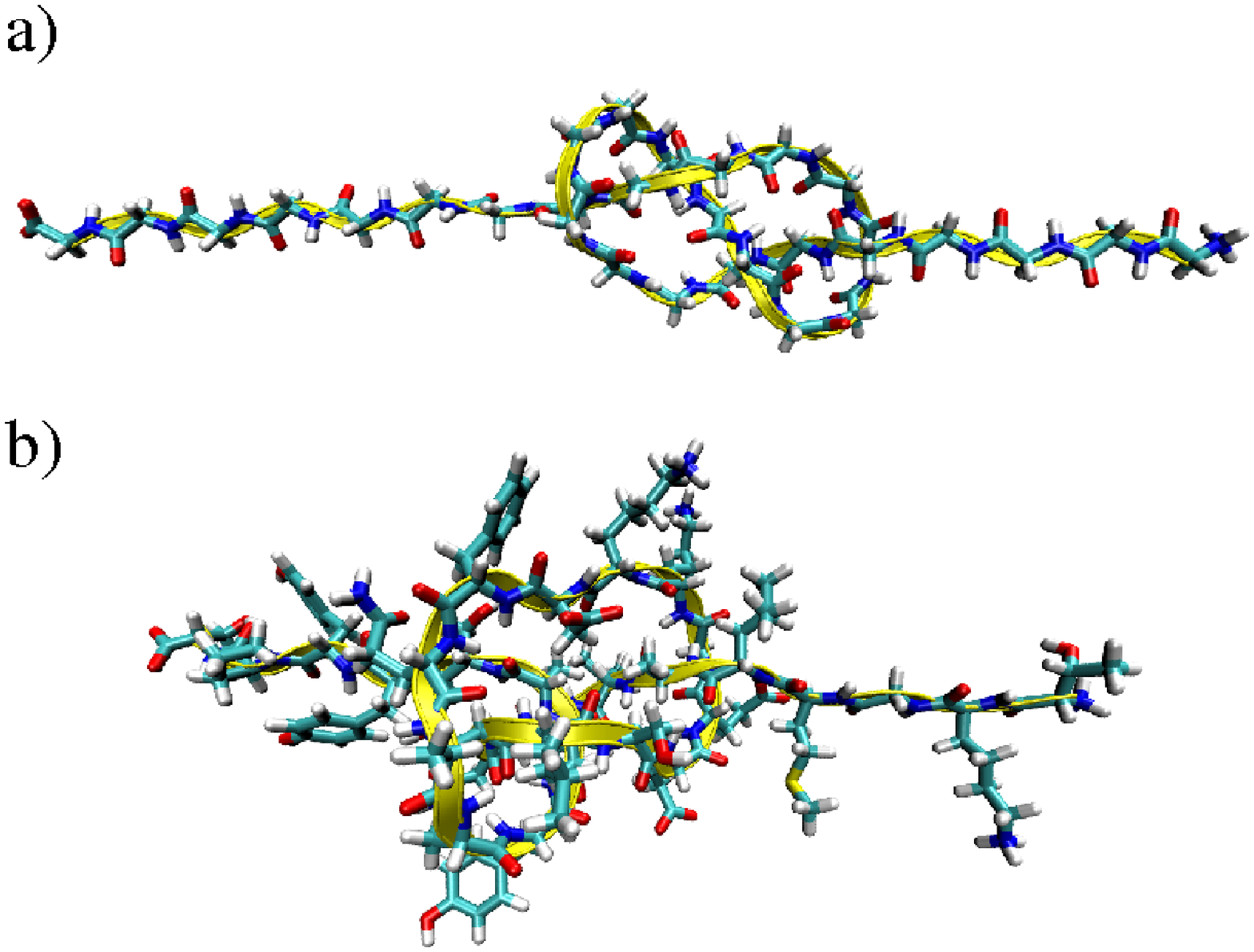}
\end{figure}
 
\newpage 

\begin{figure}[htp]
Fig. 4
\includegraphics[width=14cm,angle=0]{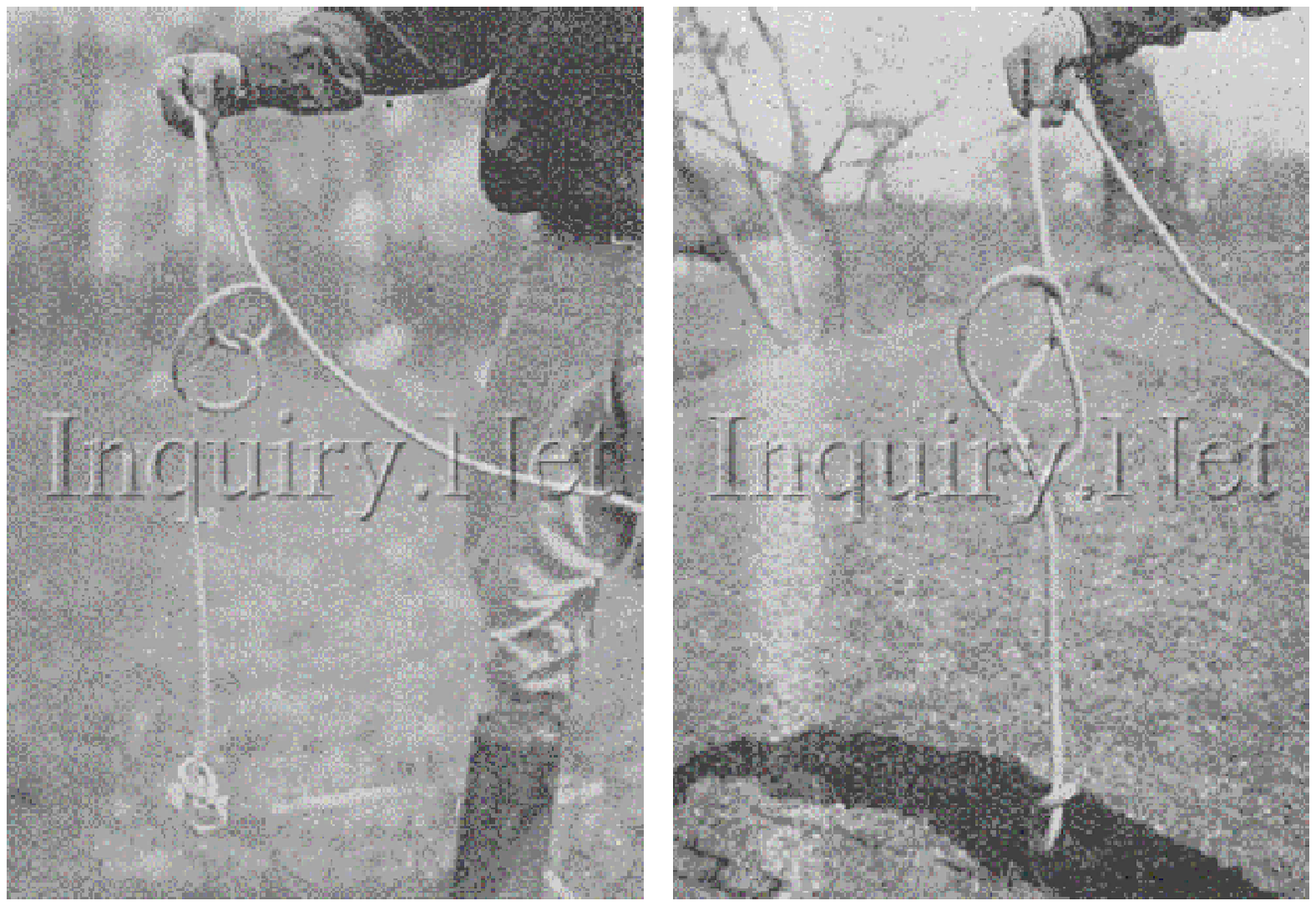}
\end{figure}

\begin{figure}[htp] 
Fig. 5
\includegraphics[width=14cm,angle=0]{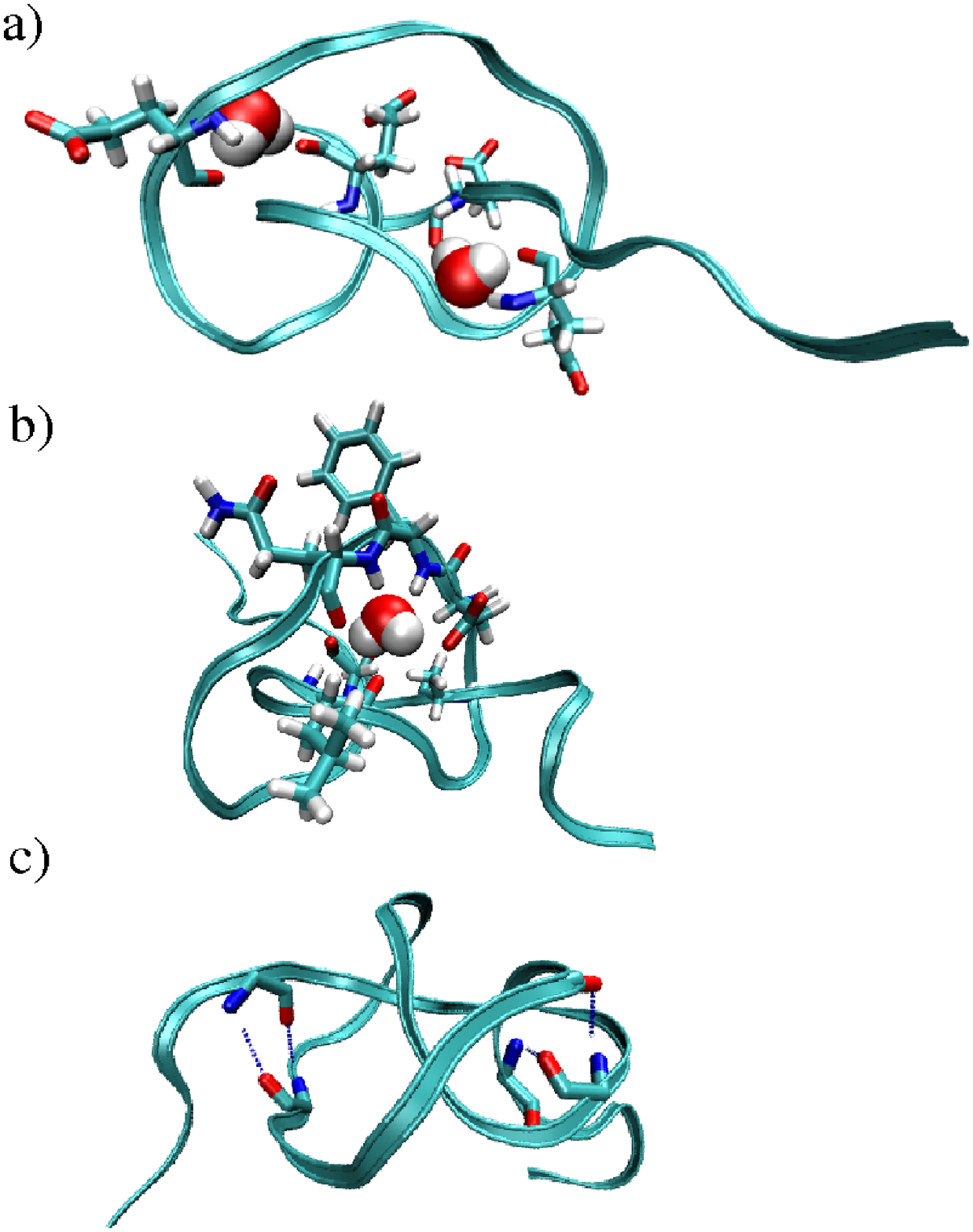}
\end{figure}


\begin{thebibliography}{53}
\bibitem{open}
\newblock Mathematically these knots are open, i.e., no closed loops, but
  throughout the work we just use the term 'knot' for simplicity.
\end{thebibliography}
\end{document}